\documentclass{llncs}
\usepackage{epsf}
\usepackage{graphicx}
\usepackage{wrapfig}
\usepackage[utf8]{inputenc}
\usepackage{enumitem}

\usepackage[disable]
	{todonotes}
\newcommand{\todoEP}[1]{\todo[color=green!40, author=\textbf{Etienne}, inline]{#1}}

\newcommand{\todoVL}[1]{\todo[color=red!40, author=\textbf{Vincent}, inline]{#1}}

\usepackage[bookmarks=true, bookmarksnumbered=true, bookmarksopen=true,
	unicode=true, colorlinks=true,
	linkcolor=blue,citecolor=blue,filecolor=blue,urlcolor=blue,
	pdfstartview=FitH
]{hyperref}

\begin{document}

\title{Impact~Of~Content~Features For Automatic~Online~Abuse~Detection}

\author{
 Etienne Papegnies, Vincent Labatut, Richard Dufour, Georges Linar\`es
}
\institute{
 LIA - University of Avignon (France)\\firstname.lastname@univ-avignon.fr
}

\maketitle

\begin{abstract}
Online communities have gained considerable importance in recent years due to the increasing number of people connected to the Internet. Moderating user content in online communities is mainly performed manually, and reducing the workload through automatic methods is of great financial interest for community maintainers. Often, the industry uses basic approaches such as bad words filtering and regular expression matching to assist the moderators. In this article, we consider the task of automatically determining if a message is abusive. This task is complex since messages are written in a non-standardized way, including spelling errors, abbreviations, community-specific codes... 
First, we evaluate the system that we propose using standard features of online messages. Then, we evaluate the impact of the addition of pre-processing strategies, as well as original specific features developed for the community of an online in-browser strategy game.
We finally propose to analyze the usefulness of this wide range of features using feature selection. This work can lead to two possible applications: 1) automatically flag potentially abusive messages to draw the moderator's attention on a narrow subset of messages ; and 2) fully automate the moderation process by deciding whether a message is abusive without any human intervention.

\end{abstract}

\section{Introduction}
\label{sec:Introduction}
On the main achievements of the Internet is the freedom and anonymity it brought into the way we communicate. Online communities, which are freely accessible exchange spaces on the Internet, have enjoyed a surge of users as a result. They come in many shapes and forms but they all share a common aspect: they have to be maintained by some party. Some online communities have gained considerable socio-economical importance due to their huge user base. A correct behavior in these communities is usually required to comply with a given set of rules so that users may enjoy a hospitable and productive environment. However, freedom and anonymity often give rise to \textit{abusive behaviors}. 
The definition of an abusive behavior is often dependent on community rules. Almost always though, users have to show respect to one another in the way they interact, so verbal abuse as well as the expression of racist, homophobic and otherwise discriminatory views constitutes abusive behaviors. 
As a result, \textit{moderation} is the task of responding to abusive behaviors by sanctioning the users exceeding the rules. This moderation work is mainly done manually, which makes it very costly in terms of human and financial costs.

In this paper, we consider the problem of automatically determining if a message from a user is abusive or not. We first present our automatic abusive message classification system based on basic features. We then propose to enrich our system by considering original preprocessing approaches, as well as corpus selection and various new content features specific to the targeted community. We finally propose a qualitative study that helps to analyze the impact of each content feature on the automatic abusive message classification performance. Two types of messages are considered in this paper: one source akin to email and the other to discussions in a chatroom, both coming from a corpus of messages originating from the community of the French massively multiplayer on-line game \href{https://play.spaceorigin.fr/}{SpaceOrigin}.

This paper is organized as follows. In Section \ref{sec:Relwork}, we review the main works related to abuse detection and automatic moderation. In Section \ref{sec:Features}, we describe the features used to automatically detect abusive messages. In Section \ref{sec:Experiments}, we present our dataset as well as the experimental protocol, and discuss the results obtained. Finally, the main points of our work are summarized in Section \ref{sec:conclusion}, which also shows how it could be extended.

\section{Related Work}
\label{sec:Relwork}
A number of works have tackled the problem of detecting abusive messages in on-line communities. While most of them are evaluated on English datasets, the majority of the methods used are language- and community-independent, and can therefore be applied on messages from any online community, which makes them relevant to us.
This review is focused around two axes: Preprocessing techniques and Features for abuse detection. {\it Preprocessing} consists in taking the raw message text and attempting to alleviate most of the problems introduced by the Internet medium, such as typos, abbreviations, use of smileys and so on. The feature extraction process consists in processing a series of indicators from the raw message text, that will reflect its class.

\subsection{Preprocessing step}
\label{sec:Preprocess}
Preprocessing is usually a simple but important step when dealing with messages posted on-line. The Internet medium introduces specific difficulties: disregard of syntax and grammar rules, out-of-vocabulary words, heavy use of abbreviations, typos, presence of URLs... The \textit{Denoising} and \textit{Deobfuscation} tasks both consist in mapping an unknown word back into a dictionary of known words. In the first case, a word is out of the vocabulary for unintentional reasons such as typos, e.g. "I uess so" for "I guess so". In the second case, it is due to a more deliberate attempt to conceal the word, e.g. "F8ck3r" for "fucker". Globally, mapping the word back into the dictionary increases the performance of probabilistic learning methods, since these methods need the cleanest possible text to achieve their maximum performance.

In \cite{sood2012using}, the \textit{Levenshtein distance} (a type of \textit{edit distance}) is used in an attempt to match unknown words against words in a crowd-sourced list of manually annotated messages containing profanity. The Levenshtein distance \cite{levenshtein1966binary} measures the number of insertions, deletions and replacements needed to convert a string into another ({\it e.g.} The Levenshtein distance between "@ss" and "ass" is 1). Computing the Levenshtein distance between two words of length $n$ and $m$ is computationally expensive: the runtime is $\mathcal{O}(nm)$, and for this specific task each word has to be matched against each word in the dictionary, which is huge. We base some of our features on the Levenshtein Distance.

Other works, such as \cite{brill2000improved}, proposed to improve on simple Levenshtein distance based denoising (for the purpose of spell checking) by considering the context of the string edition in a word (where the edited character is) and in a sentence (does the edit maps the word into a high $n$-gram probability?). However, this approach is based on the assumption that out-of-vocabulary words are mainly due to unintentional spelling mistakes and is therefore not applicable to deobfuscation.

In \cite{lee2005spam}, the authors use a Hidden Markov Model customized with dictionary and context awareness for the purpose of deobfuscating spam messages. Those types of messages often include deliberately misspelled words in an attempt to bypass filters. Their model showed impressive results with the ability to correct both unintentional misspellings as well as deliberate obfuscation using weird characters or digits and could even map segmented words back into complete words. (ie: "ree movee" $\rightarrow$ "remove"). This preprocessing approach, while effective, is computationally intensive and complex to implement.

Preprocessing is an important step in an automated abuse detection framework, but it should be applied with caution. The goal of preprocessing is to increase the amount of relevant information in a message, but it can have the opposite effect. For example the tendency of a user to misspell words can be viewed as an important feature to describe the user, but blind preprocessing would hide that.

\subsection{Text messaging classification}
\label{sec:ContentContextBased}
In this section we review existing classification approaches that consider the content of the messages to detect abusive messages, and then the context of the exchanged messages.

\paragraph{Content-based approaches.}
The work described in \cite{spertus1997smokey} was one of the first to automate the detection of hostility in messages. While hostility does not imply abuse, abuse often contains hostility. The paper defined a number of rules to identify certain characteristics of the messages, such as Imperative Statement, Profanity, Condescension, Insult, Politeness and Praise. A Decision Tree classifier was then used to categorize messages into Hostile and Non-hostile classes. The setup showed good results but was limited when dealing with sarcasm, noise or innuendo. This approach is interesting but highly tuned for the grammar rules, semantics and idioms of a specific language. 

In \cite{chen2012detecting}, the authors note that the mere presence of an offensive word in a message is not a strong enough indication that the message itself is offensive, i.e. "You are stupid." is way more offensive than "This is stupid.". This is an important observation and the authors showed that the lack of context can be mitigated by looking at word $n$-grams instead of unigrams.

Another work \cite{dinakar2011modeling} used features computed from $tf$-$idf$ weights, a list of words reflecting negative sentiment and widely used sentences containing verbal abuse to detect cyberbullying in comments associated with Youtube videos. Their model showed good results for instances of verbal abuse and profanity but was limited with regard to sarcasm and euphemism. 

Finally, in \cite{chavan2015machine} the authors reviewed machine learning approaches for the classification of aggressive messages in On-line Social Networks, described a full pipeline for achieving classification of raw comments and introduced two new features: Pronoun Occurrence and Skip-Gram features. They allow for the detection of targeted phrases such as "He sucks" or "You can go die", and for the identification of long distance relationship between words, respectively.

Content-based text classification performs relatively well as a starting point. The computational cost to implement these approaches is usually reasonable. Nonetheless, these approaches have severe limitations. For instance, abuse can cross message boundaries, and therefore a message can be abusive only because of the presence of an earlier message. In other cases, messages can be abusive because they reference a shared history between two users. Therefore, studying the context of a message, its recipient, and its author might also be important.

\paragraph{Context-based approaches.}
To go beyond the limitations of content-based approaches, some authors proposed to take into account the context of messages, usually in addition to the textual content itself.

Some works explore the use of the content neighboring the targeted message. In \cite{yin2009detection}, the authors used a supervised classifier working on $n$-gram features, sentence-level regular expression patterns and the features derived from the neighboring phrases of a given message to detect abuse on the Web. Their approach focused on detecting derivations in the context of a discussion around a given topic and their context features significantly improved the performance of their system. For this reason, we want to adopt a similar approach for our own method, but by focusing on derivations of users themselves from their usual patterns.

Other works have focused on modeling the users' behaviors by introducing higher-level features than the textual context. A comprehensive study of antisocial behavior in on-line discussion communities has been proposed in \cite{cheng2015antisocial}. Their exploratory work reinforced the weight of classic features used to classify messages such as misspellings and length of words, and provided insight into the devolution of users over time in a community, regarding both the quality of their contributions and their reactions towards other members of the community. This analysis is a good step towards modeling abusive behavior. One of the essential results of the analysis is that instances of antisocial messages tend to generate a bigger response from the community compared to benign messages. The number of respondents to a given message is a feature we use in this work.

A selection of contextual features aiming at detecting abuse in on-line game communities has also been investigated in \cite{balci2015automatic}. These features form a model of the users of the game by including information such as their gender, number of friends, investment in the platform, avatars, and general rankings. The goal was to help human moderators dealing with abuse reports, and the approach yielded sufficiently good results to achieve this goal. The work was however limited in applicability because of the specifics of that given community, and of the raw amount of data needed to perform similar experiments in other games. 

When quantifying controversy in social media \cite{garimella2016quantifying}, the structure of the community network is exploited to identify topics that are likely to trigger strong disagreements between users. The approach relies on a network whose nodes are Twitter users and links represent communications between them. It is interesting, however hardly applicable to our case, since we cannot infer the exact network structure from our dataset unless we restrict the network to private conversations between two users.

\section{Abusive Message Features} 
\label{sec:Features}
In this section, we describe the content-based features used in our automatic abusive message classification system. They can be broadly categorized as morphological, language, and context features. Some of them are quite generic, in the sense they are used for different classification tasks in the literature. The others were designed by us specifically for this experiment, and some are customized for the community where our dataset originates, but they can sometimes be generalized to other communities (we reflect on this in Section \ref{sec:conclusion}). The features we developed are denoted with a star (*) preceding their name and description.

Some features require the data to be preprocessed before being extracted, so we start with the description of our preprocessing approach first.

\subsection{Preprocessing}
\label{sec:Preprocessing}
We distinguish two preprocessing phases. In the \textit{basic phase}, we first lower-case the raw text and tokenize it using spaces. Each token in the list is then stripped of punctuation before the message is reassembled.

In the \textit{advanced phase}, the data undergo some additional preparation steps. First, we revert elision. \textit{Elision} refers to the suppression of a final unstressed vowel immediately before another word beginning with a vowel. For the French language, we therefore replace instances of "j'", "t'" by their respective long forms "je ", "te ", so that, for instance, "j'arrive" becomes "je arrive". Second, we run a deobfuscation pass by mapping hexadecimal or binary encoded text in the message back to ASCII. This is highly specific to the considered online community, because users sometimes encode part of their messages in that way. Third, we convert each URL into a sequence of tokens. The first describes whether this URL is an internal link (to a server hosting the community) or an external one. The rest are words that could possibly be extracted from the name of the web page. For instance: \texttt{http://edition.cnn.com/2017/01/31/politics/donald\-trump-immigration-white-house/index.html} is mapped to: \textbf{\_\_url\_external cnn com politics donald trump immigration white house index}. Finally, we use the FrenchStemmer from the Natural Language Toolkit~\cite{bird2006nltk} to convert words into their stem.

\subsection{Morphological Features}
\label{sec:MorphoFeats}
\paragraph{Message Length.} 
This feature corresponds to the length of a message, expressed in number of characters, before any pre-processing. The intuition is that abusive messages are usually either kept short ({\it e.g.} "Go die."), or extremely long, which is symptomatic of a massive copy/paste.

We also consider the length expressed in terms of words. In conjunction with the character count, it can match certain overly emphasized messages (e.g.: "Shuuuuuuuuuuuuuuuuuuuuuuut uuuuuuuuuuuuuuuuuuuup!").

\paragraph{Character Classes.} 
We split characters into $5$ classes: \textit{Letters}, \textit{Digits}, \textit{Punctuation}, \textit{Spaces} and \textit{Others}. We keep track of the number of characters in those classes and the ratio of those classes in the message. This is done on the raw message, before any preprocessing.

We selected these features based on several observations we made on the abusive messages. First, some of them have an unusual number of characters in the "Other" class, e.g.: "8==================D". Second, some use digits to obfuscate their meaning, in violation of the game rules. For instance, "01000111011011110010000001100100011010010110010100101110" and "476F20\-6469652E" are obfuscated versions of the text "Go die.": the first one is coded in binary, and the second in hexadecimal.

Abusive users also commonly "yell" insults using capital letters, which is why we keep track of both the number of caps and the corresponding ratio of caps in the message.

\paragraph{Compression Ratio.} 
This feature is defined as the ratio of the length of the compressed message to that of the original message, both expressed in characters. It is based on the observation that certain users tend to repeat \textit{exactly} and many times the same text in their abusive messages. We use the Lempel–Ziv–Welch (LZW) compression algorithm~\cite{batista2004texture}, and the feature therefore directly relates to the number of copy/pastes made in the same message.

\paragraph{Unique Characters.} 
By counting the number of distinct characters in the message, we can detect the use of binary or hexadecimal obfuscation, as well as the overuse of punctuation. For instance, for the message "010001110110111101110", this feature has only a value of $2$. This value is also computed before any preprocessing.

\paragraph{Collapsed Characters.} 
This feature is computed after the message is lowercased. When three or more identical consecutive characters are found in the message, they are collapsed down to two characters. For instance, "looooooool" would be collapsed to "lool". The feature is the difference between the length of the original message and that of the collapsed one. This preprocessing step has been widely used in the classification of Tweets, for instance in \cite{roy2013lexicon}.

\subsection{Language Features}
\label{sec:LangFeats}

\paragraph{Bag of Words.} We transform the message into a Bag of Words (BoW). This is a sparse binary vector that has the same dimension as the known vocabulary of the corpus and where each component corresponds to one word. The component has value 1 if the word is present in the message and 0 if it is not. We use the output of a Naive Bayes classifier for a given BoW as an input feature into our larger system.

\paragraph{Word Length.} 
This feature is a component of the Automated Readability Index (ARI) \cite{Senter1967}. It measures how proficient someone is at creating text documents. While abusive messages are sometimes surprisingly well written, this remains rare.

\paragraph{Unique Words.} 
We consider the number of unique words in the message. The intuition is that messages with more words are likely to be more constructive in terms of their content. Moreover, we observed that people are generally straightforward when verbally abusing others, and rarely take the time to elaborate.

\paragraph{$tf$-$idf$ Scores.} 
Those features are the sums of the $tf$-$idf$ scores of each individual word in the message. We use two distinct scores: the so-called \textit{non-abuse score} is processed relatively to the non-abuse class (randomly chosen messages that have not been flagged as abusive), whereas the \textit{abuse score} is processed over the abuse class.

If the considered word is unknown, in the sense it does not appear in the training set, we process an approximation of these scores. For this purpose, we first search for known words located within a given Levenshtein distance from the word of interest, and average their own scores.

Computing the full Levenshtein Distance between two words is computationally expensive. For this reason, solutions proposed in this paper never compute the full Levenshtein Distance between two words. Instead, a specialized tree-based index data-structure with a search function that yields all words in the tree within a given maximum edit distance is used. We use 2 as the maximum edit distance. This is considerably faster because branches of the tree are pruned as soon as we reach a state where the maximum edit distance is exceeded. It is still the second most computationally intensive operation in our experiments.

\paragraph{Sentiment Scores.} 
These features are numeric values derived from the number of words in the message that have a sentiment weight. It is based on the sentiment corpus presented in \cite{P14-2063}, which was automatically generated for the French language. We augmented it manually by selecting words from a large list of insults.

\paragraph{Bad Words.} 
This feature corresponds to the number of words in the message that appear in a manually crafted list of bad words. The list of words was created from a list of insults in French and then augmented with common Internet shorthand and symbolism. (\textit{e.g}: 'connard', 'fdp', 'stfu' but also '..|..', '8==D' etc)

When we cannot match the considered word to any of the bad words in our list, we try to perform a fuzzy match using the Levenshtein distance. This is supposed to allow us picking up some obfuscated bad words.

We also perform the same tests using the collapsed version of the message (as described for the \textit{Collapsed Characters} feature).

\paragraph{Business Score*.}
We first mine messages for patterns specific to the community. In the targeted context of this particular online strategy game, we chose to focus on: names of buildings and military units, war vocabulary and other game-specific jargon, sets of Coordinates, and report links. The latter refers to internal URLs generated by some actions, and pointing towards summaries that the players can share.
For each pattern, we manually developed a regular expression and used it to find the number of non-overlapping occurrences of the pattern in the message. We then produced the Business Score by combining the individual scores obtained for each pattern. This is a measure of how the message relates to the focus of the community. By observing the corpus we noticed that abusive messages tend to be strictly personal attacks with no pretense of roleplay and no mention of game jargon.

\subsection{User Behavior Features}
\label{sec:UserFeats}

\paragraph{Number of respondents.} 
Given a fixed size window after a target message, this feature tracks how many distinct users replied to this message. This feature is likely to be relevant, because it has been shown that abusive comments tend to trigger big responses from the community.

\paragraph{Probability of $n$-gram emission (PNE)*.} 
We investigate if the abusiveness of a user's message can be detected by considering the effect it has on the other users participating to the same conversation. To do this, we compare the writing behavior of the other users before and after the apparition of the targeted message. We model this behavior through a user-specific Markov chain, which we use to compute how likely some text is to have been generated by the considered user.

We first sort the messages in chronological order. For each participant other than the user who wrote the targeted message, we build a word $n$-gram Markov chain using all but the last $W$ $n$-grams in the messages posted before the target message.

\begin{figure}
	\centering
	\vspace{-0.5cm}
	\includegraphics[width=\textwidth]{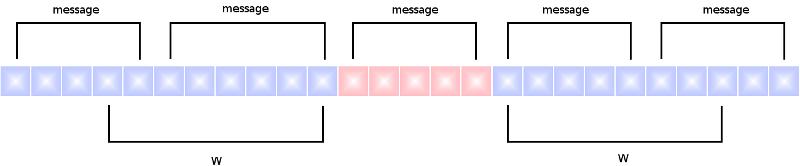}
	\vspace{-0.5cm}
	\caption{A sequence of messages broken down into $n$-grams. Each square represents an $n$-gram: red for the targeted message, blue for the surrounding messages written by other users.}
	\label{fig:markov}
\end{figure}

The Markov chain is a convenient way to store the transition probabilities for all couples of $n$-grams in a participant's history. We compute two values: the average emission probabilities of the $n$-grams in the $W$-length window before and after the target message, as represented in Figure \ref{fig:markov}.

Let $P_{i,i+1}$ be the emission probability of a transition between the $i^{th}$ and $i+1^{th}$ $n$-grams in the window of length $W$. Then we define the average emission probability $S$ over the set of $W$ $n$-grams as:
\begin{equation}
	S = \frac{\sum_{i=0}^{W-1}{P_{i,i+1}}}{W}
\end{equation}

We note $S_B(u)$ and $S_A(u)$, respectively, the average probabilities processed before and after the targeted message, for the same user $u$. The final score $S(u)$ for user $u$ corresponds to their difference:
\begin{equation}
	S(u) = S_A(u) - S_B(u)
\end{equation}

This score is processed for every respondent to a message in a window of fixed length after the message. We then compute our feature by averaging this score over all the responding users.

\paragraph{Applicability Criterion for PNE*.} 
The previous feature requires averaging scores, which makes sense only if the considered users have sufficient history: we define a limit of at least $300$ bigrams. This feature reflects the fulfillment of this constraint.

\section{Experiments} 
\label{sec:Experiments}
In this section, we describe the data used in our experiments (Subsection~\ref{sec:Dataset}) as well as the experimental protocol (Subsection~\ref{sec:Setup}). We then evaluate the proposed system, including the various features and original preprocessing approaches (Subsection~\ref{sec:Results}).

\subsection{Dataset} 
\label{sec:Dataset}
We have access to a database of users' in-game interactions for the considered MMO. This user-generated content was manually verified, in the sense the game users had the ability to flag parts of the content as inappropriate ({\it i.e. abusive}). There are many types of reportable contents, but, in this paper, we focus on two of them: ingame-messages (iM) and chat messages (cM), collectively referred to as messages.

\textit{Ingame-messages (iM)} are on-line messages with a clearly defined reach. They are the equivalent of e-mails and can be sent to specific users or groups of users. They can be edited \textit{a posteriori} by moderators when an abuse case is reported. The reach of \textit{chat messages (cM)} is loosely defined because it is limited to users currently active in a chatroom. However, there is no way to determine which user has actually seen a specific chat-message based on the available data. Users are fed recent scroll history for a chatroom upon joining, but it is not possible to reliably determine who has joined when from the chat logs. Chat-messages cannot be edited by moderators afterwards.

The database contains $474,599$ in-game messages and $3,554,744$ chat-messages. We extract $779$ \textit{abusive} messages ($0.02\%$), which constitute what we call the \textit{Positive Class} (Class 1) of messages. These messages were first flagged by the game users using a built-in reporting tool, and then confirmed as being abuse cases by the game community moderators. Of these $779$ abusive messages, $14\%$ are ingame-messages, and the rest are chat-messages.  We then extract \textit{non-abuse} messages from the database, in order to constitute the so-called \textit{Negative Class} (Class 0). They are chosen at random from a pool of messages written by users which have \textit{never} been flagged by a confirmed abuse report. For each message, we also gather context data: a window of messages occurring before and after each message.

We run the experiments with different versions of the corpus: in-game messages only (\textit{iM}), chat-messages only (\textit{cM}) and messages of both types combined (\textit{iM+cM}). Sizes of each considered corpus configuration are reported in Table~\ref{tab:data}. These configurations are considered as ``unbalanced'' (U), since there are twice as many non-abusive messages as abusive messages. As a result, we also experiment with the use of ``balanced'' data (B), where the number of abusive messages is equals to the non-abusive ones.

\begin{table}[htp]
	\center
	\vspace{-0.5cm}
	\begin{tabular}{l@{\hskip 0.5cm}r@{\hskip 0.5cm}r}
		\hline
		\textbf{Configuration} & \textbf{Abusive Messages} & \textbf{Non-Abusive Messages} \\ 
        \hline
		iM+cM & 779 & 1558 \\
		iM & 111 & 222 \\
		cM & 668 & 1336 \\
		\hline
	\end{tabular}
	\vspace{0.2cm}
	\caption{Corpus sizes depending on the considered experimental setup (unbalanced data).}
	\label{tab:data}
	\vspace{-0.5cm}
\end{table}

\subsection{Experimental Setup}
\label{sec:Setup}

\begin{figure}[!ht]
	\center
	\vspace{-0.5cm}
	\includegraphics[width=0.7\textwidth]{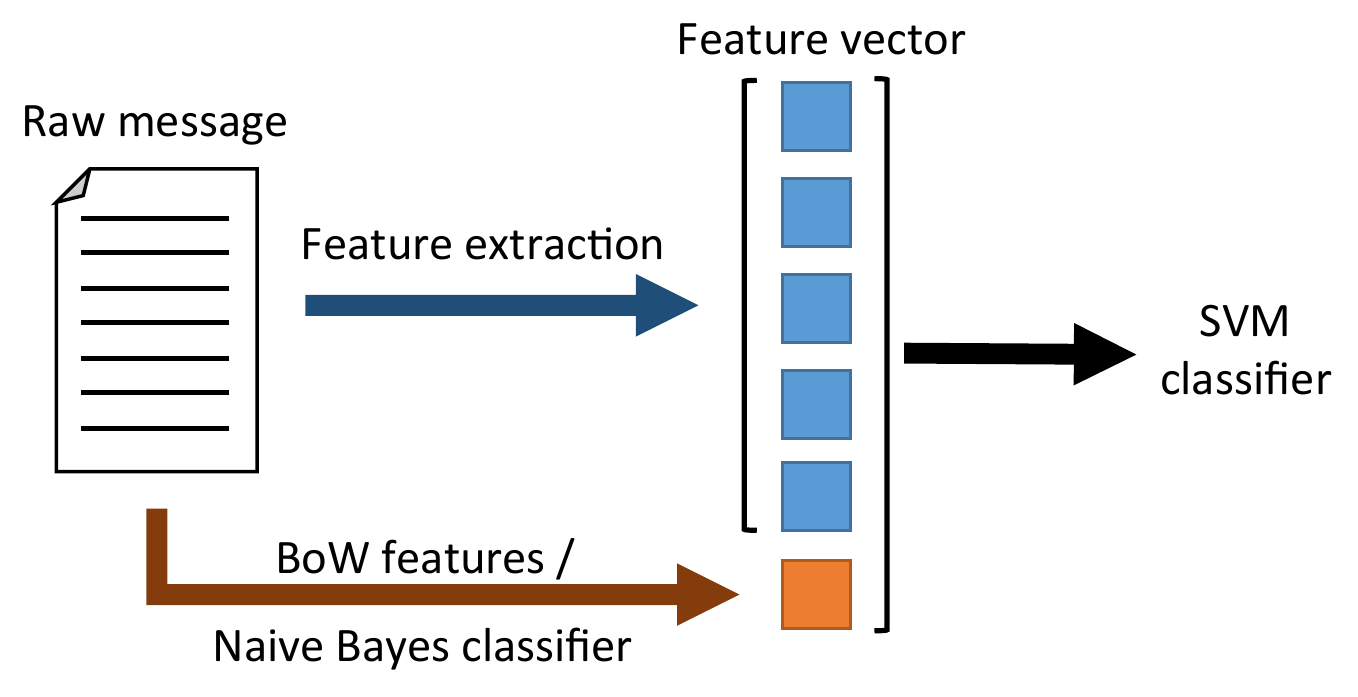}
	\caption{The full experimental setup}
	\vspace{-0.5cm}
	\label{fig:Pipeline}
\end{figure}

Our experiment is designed as a multi-stage classifier pipeline, as described in Figure \ref{fig:Pipeline} (each box corresponds to a stage). The first stage (\textit{Raw Messages}) consists in building the corpus. Messages from both the Abuse and Non-abuse classes are extracted from the database as explained in Subsection~\ref{sec:Dataset}. The corpus is then split into a \textit{Train} set containing $70\%$ of the messages, and a \textit{Test} set containing the remaining $30\%$.

In the second stage (\textit{Bag of Words Features}), messages are normalized, tokenized and converted into Bag Of Words. In the third stage (\textit{Naive Bayes Classifier}), the Bag Of Words representations of the \textit{Train} messages are used to build a Naive Bayes classifier. This classifier is then used to generate predictions for the class of the \textit{Test} messages.

In the fourth stage (\textit{Feature Extraction}), we extract the features described in Section \ref{sec:Features} from the messages. As explained before, some of these are derived from the messages before any normalization or preprocessing, whereas some others require a specific preprocessing. We then use another classifier, a Support Vector Machine (SVM). We could directly feed the Bag Of Words to the SVM. However, given the size of the vocabulary in our experiments, this would lead to a dimensionality issue, with a number of features greatly exceeding the number of instances in the corpus. Therefore, we prefer to consider the decision from the Naive Bayes classifier (third stage) as an additional feature given to the SVM. We get a total of $67$ distinct features, including the Naive Bayes decision, which are all gathered into an array.

The fifth stage (\textit{SVM Classifier}) is the final classification: the feature arrays from the Train set are fetched to an SVM classifier, and the resulting model is then used to generate class predictions for the Test set.

\subsection{Results}
\label{sec:Results}

We evaluate the performance of our proposed abusive message detection system in terms of the traditional Recall, Precision and $F$-Measure. Given the relatively low number of abusive samples of the targeted corpus, the whole dataset was split into $10$ parts and every result given in this section is the average value over a $10$-fold cross validation. In order to show the contribution of the features as well as pre-processing approach proposed, three system configurations are studied. The first is the \textit{baseline}, which relies on the classic feature set and the basic preprocessing, as previously described. The two others are our contributions: on the one hand the full feature set with basic preprocessing, and on the other hand the full feature set with advanced preprocessing.

Table~\ref{tableAllUnbalanced} presents the performance obtained by the proposed system for all the studied configurations, using \textit{unbalanced} data. We can firstly see that, no matter the considered message type (iM only, cM only, or iM+cM), improvements in terms of Precision, Recall and $F$-measure are observed when completing the baseline system (classic features and basic preprocessing) with our new features. This gain is even more important when using our advanced preprocessing, with $F$-measure increases of $3.1$ points (iM only), $3.3$ points (cM only) and $3.2$ points (iM+cM) compared to the baseline system. The same observations can be made for the results obtained on the \textit{balanced} data, displayed in Table~\ref{tableAllBalanced}, but with smaller gains ($3.3$, $1.4$ and $1.3$ points, respectively).

\todoEP{Update results following $class\_weigth='balanced'$ fix}
\todoEP{Progress: All results up to date}
\todoEP{CHeck result discussion if numbers needs to be changed}
\begin{table}[htp]
	\center
	\vspace{-0.5cm}
    \begin{tabular}{ l@{\hskip 0.5cm}l@{\hskip 0.5cm}l@{\hskip 0.5cm}r@{\hskip 0.5cm}r@{\hskip 0.5cm}r }
		\hline
    	\textbf{Data} & \textbf{Features} & \textbf{Preprocessing} & \textbf{Precision} & \textbf{Recall} & \textbf{$F$-Measure} \\ 
        \hline
        \textbf{iM only} & \textbf{Classic set} & \textbf{Basic} & 66.9 & 72.8 & 69.7 \\  
         & \textbf{Full set} & \textbf{Basic} & 67.2 & 73.4 & 70.2 \\ 
         & \textbf{Full set} & \textbf{Advanced} & {\bf 69.6} & {\bf 76.2} & {\bf 72.8} \\ 
         \hline
        \textbf{cM only} & \textbf{Classic set} & \textbf{Basic} & 65.2 & 71.6 & 68.2 \\ 
         & \textbf{Full set} & \textbf{Basic} & 65.5 & 72.2 & 68.7 \\ 
         & \textbf{Full set} & \textbf{Advanced} & {\bf 67.6} & {\bf 75.9} & {\bf 71.5} \\ 
         \hline
        \textbf{iM+cM} & \textbf{Classic set} & \textbf{Basic} & 65.7 & 72.3 & 68.9 \\ 
         & \textbf{Full set} & \textbf{Basic} & 65.9 & 73.2 & 69.3 \\ 
         & \textbf{Full set} & \textbf{Advanced} & {\bf 68.3} & {\bf 76.4} & {\bf 72.1} \\ 
         \hline
	\end{tabular} 
	\vspace{0.2cm}
	\caption{Classification results (in \%) of the automatic abusive message classification system, obtained by applying different feature sets and preprocessing configurations to the \textit{unbalanced} data.}
	\label{tableAllUnbalanced}
	\vspace{-0.5cm}
\end{table}

Let us now compare the results obtained for the different types of messages. When considering the unbalanced data (Table~\ref{tableAllUnbalanced}), iM and cM only lead to globally similar performances for all three considered measures. Combining them (iM+cM) does not bring any significant change. However, this is not the case for the balanced data (Table~\ref{tableAllBalanced}): the performance obtained for cM only is quite different, with a much higher Precision ($+7.6$ points on the advanced setup) and a lower Recall ($-2.9$ points). This pulls up the overall performance when using both message types (iM+cM), leading to a $76.5$ $F$-Measure for the advanced setup, which is $4.4$ points higher than with the unbalanced data.

\begin{table}[htp]
	\center
	\vspace{-0.5cm}
	\begin{tabular}{ l@{\hskip 0.5cm}l@{\hskip 0.5cm}l@{\hskip 0.5cm}r@{\hskip 0.5cm}r@{\hskip 0.5cm}r }
		\hline
    	\textbf{Data} & \textbf{Features} & \textbf{Preprocessing} & \textbf{Precision} & \textbf{Recall} & \textbf{$F$-Measure} \\ 
        \hline
        \textbf{iM only} & \textbf{Classic set} & \textbf{Basic} & 67.2 & 70.4 & 68.7 \\ 
         & \textbf{Full set} & \textbf{Basic} & 67.6 & 70.8 & 69.2 \\
         & \textbf{Full set} & \textbf{Advanced} & {\bf 70.8} & {\bf 73.3} & {\bf 72.0} \\ \hline
        \textbf{cM only} & \textbf{Classic set} & \textbf{Basic} & 77.2 & 67.5 & 72.0 \\
         & \textbf{Full set} & \textbf{Basic} & 77.1 & 67.4 & 71.9 \\
         & \textbf{Full set} & \textbf{Advanced} & {\bf 76.8} & {\bf 70.3} & {\bf 73.4} \\ \hline
        \textbf{iM+cM} & \textbf{Classic set} & \textbf{Basic} & 76.9 & 73.6 & 75.2 \\ 
         & \textbf{Full set} & \textbf{Basic} & 77.3 & 74.5 & 75.9 \\ 
         & \textbf{Full set} & \textbf{Advanced} & {\bf 76.1} & {\bf 76.9} & {\bf 76.5} \\ \hline
	\end{tabular}
	\vspace{0.2cm}
	\caption{Classification results (in \%) of the automatic abusive message classification system, obtained by applying different feature sets and preprocessing configurations to the \textit{balanced} data.}
	\label{tableAllBalanced}
	\vspace{-0.5cm}
\end{table} 

\begin{figure}[htp]
	\centering
	\vspace{-0.5cm}
	\includegraphics[width=0.49\textwidth]{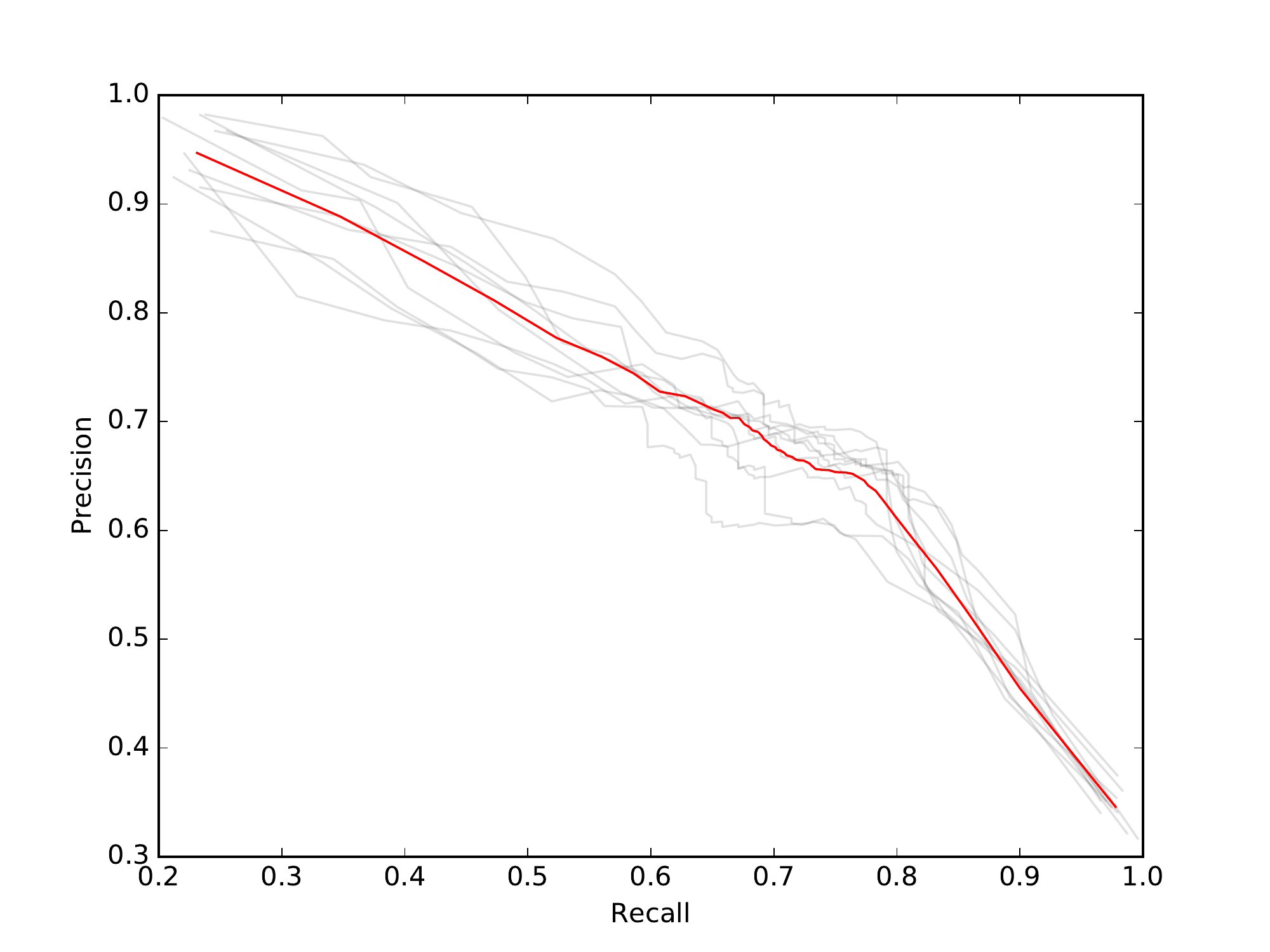}
	\includegraphics[width=0.49\textwidth]{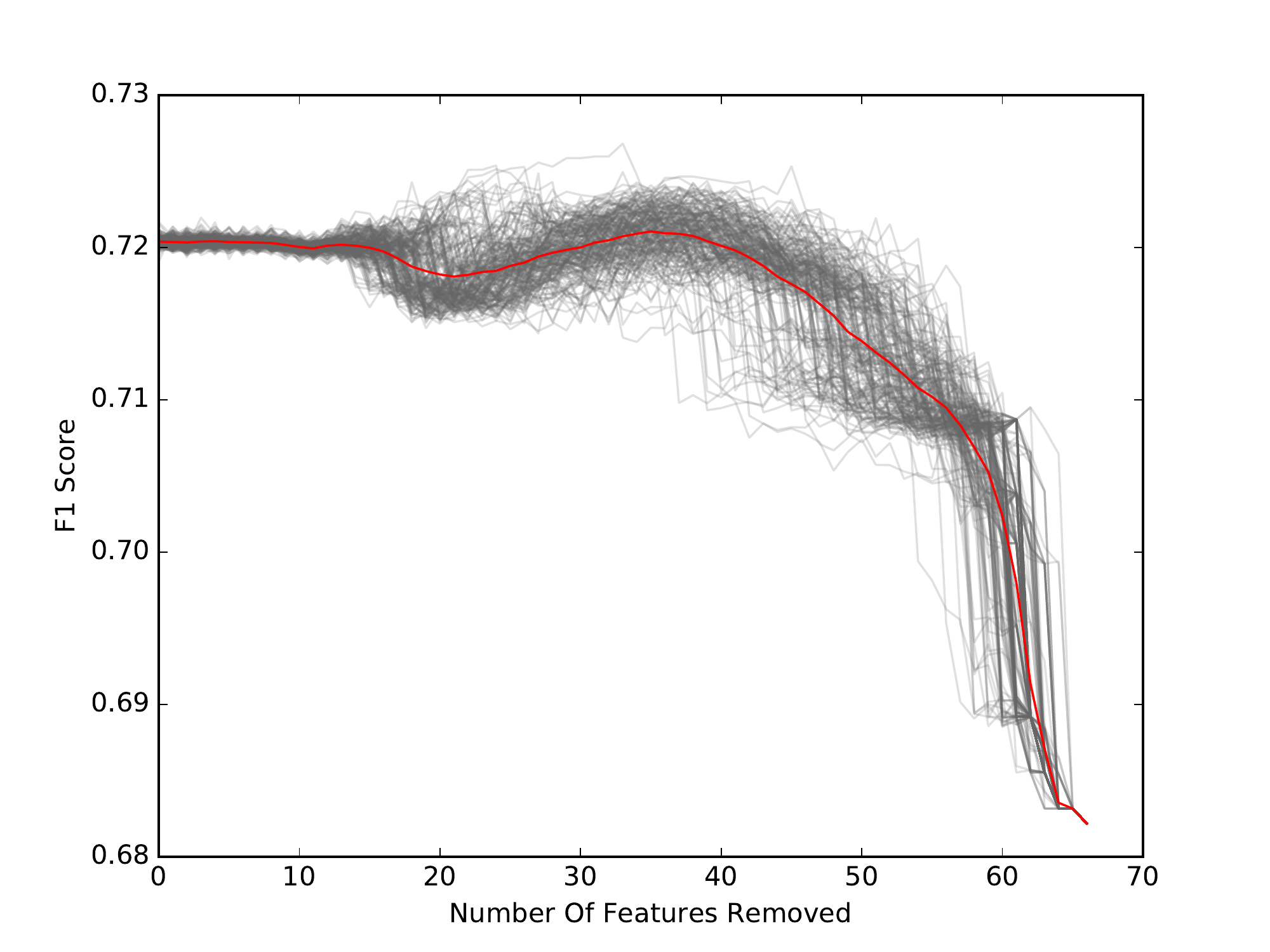}
	\vspace{-0.5cm}
	\caption{Left: Precision-Recall curve of the SVM classifier. Right: Evolution of the classifier performance when sequentially dropping all features but one.}
	\label{fig:Curve}
\end{figure}

Our experiments show that, even if acceptable results could be obtained with our abusive message detection system (best $F$-measure of more than 70\%), performance is still not good enough to be directly used as a fully automatic system that replaces human moderation. Nonetheless, we think that this system could be useful to help moderators focus on messages considered as potentially abusive, instead of having to analyze all messages. This is illustrated by the left plot in Figure~\ref{fig:Curve}, which represents the Precision-Recall curve (traditionally obtained by varying the decision threshold on the SVM posterior probability obtained by applying the Platt Scalling implementation of the Scikit-Learn Library ~\cite{pedregosa2011scikit}). For a fully automatic system, requiring to be very precise on the decision to take ({\it i.e.} be sure that the message is abusive), a higher threshold should be used, with a loss in terms of number of detected abusive messages ({\it i.e.} lower Recall). On the contrary, for a software assisting a moderator, needing to recover as many abusive messages as possible, a lower threshold should be used, resulting in a higher recall (more abusive messages are retrieved) associated to a lower precision (more non-abusive messages be wrongly returned by the system). The plot shows a short plateau in the middle, which means it would be possible to increase the Recall without losing much Precision. However, estimating the exact optimal decision threshold will require more data.
\todoEP{Updated figure on the right. Maybe discuss the slight bump?}
\todoEP{Updated PR curve w/ PRs for each classifier and average PR curve}
\todoVL{Petit probleme de consistance entre la perf affichee dans ces figures et celles des tables. Trop long a recalculer pour la version soumise, mais il faudra corriger dans la suivante.}
We now take a look at how the features are contributing to the result. We use a tree-based estimator from the Scikit-Learn library to estimate the importance of the features for our classification problem. This tool is stochastic, so the score measuring this importance can vary from one run to the other. Thus, we ran it $200$ times to get stable results. The right plot in Figure~\ref{fig:Curve} shows all of these runs as well as the average curve.  It displays how the $F$-Measure evolves as the features are removed one by one, by increasing order of estimated importance. Our SVM classifier is trained and evaluated at each feature removal. Despite the stochastic nature of the process, the last removed (and therefore most important) feature is always the Naive Bayes decision: this makes sense, since it is already the output of a full-fledged classifier. This is confirmed by the tree-based estimator, which gives an importance score of $42.5\%$.

Each of the $200$ runs shows a sharp drop at the end. We detected that this drop is due to the removal of any feature in the following group: Number of bad words in the collapsed comment, Average word length, PNE and Applicability criterion for PNE. We therefore conclude that these features are complementary, and result in a strong classifier when combined. According to the tree-based estimator, these four features have a combined importance score of $15.9\%$. So, our results show that a small group of $5$ features account for $58.4\%$ of the classifier performance. The rest of the features improve the performance only marginally. Other noteworthy features include the ratio of letters and other characters ($5\%$), the ratio of capitalized letters ($2.1\%$), and the positive and negative scores ($4.23\%$). The Business Score feature, defined by us specifically for the targeted online community, has only an importance of $1.13\%$: it accounts for a small part of the classifier decision, but on the positive side it is fast to compute. This is not the case for the PNE feature: computing it is expensive both in terms of CPU time and memory since we need to build and store a complete model of multiple user speech patterns.

\section{Conclusion}
\label{sec:conclusion}
In this paper, we developed a system to classify abusive messages from an on-line community. It is developed on top of a first-stage Naive Bayes classifier and relies on multiple types of features: morphological, language- and context-based features, that have proven their usefulness in previous research. We added several features that we derived directly from observations of our corpus, and developed a context-based feature that aims to capture abnormal reactions from users caused by an abusive message. Our goal here was to explore a large number of features to identify the most relevant one for the problem at hand.

Our results show that abusive messages have characteristics that can be caught by an automatic system, our proposed system achieving a Recall and a Precision of more than $76\%$ on our dataset. While the performance of the system is not good enough yet to be deployed as fully automatic moderation tool, this can already help moderators focus on messages being identified as abusive, before a manual verification is made. However, because some features used in the system are specific to the community in which it is meant to operate, care must be taken when adapting the system to work on a different dataset. Our results also show that a small number of features, including both generic and problem-specific ones, account for most of the classifier decision. 

We now plan to pursue our work in several ways. First, because preprocessing has been shown to have an important effect on overall performance, we will experiment with computationally more demanding preprocessing methods, such as the HMM-based preprocessor from \cite{lee2005spam}, and evaluate their contribution to the classifier performance. Second, we want to derive variants of our PNE feature, and assess which one is the most appropriate in our situation. More generally, we plan to propose other context-based features, especially ones based on the network of user interactions.

\bibliographystyle{splncs}

\bibliography{cycling}

\end{document}